\magnification=\magstep2

\hsize=6.45truein
\vsize=8.89truein
\hoffset=-0.43truein
\voffset=0.01truein

\pageno 1

\baselineskip=14.9pt

\def\vecS{{\vec S}}
\def\vecs{{\vec s}}

\def\lsim{\,$\raise0.3ex\hbox{$<$}\llap{\lower0.8ex\hbox{$\sim$}}$\,}

\noindent
\centerline{\bf Numerical Study of the Spin-1 Antiferromagnetic}

\centerline{\bf Heisenberg Chain with a Spin-1/2 Impurity}

\vskip 12pt

\centerline{Takashi T{\sevenrm ONEGAWA}\footnote{$^{\ddagger}$}{tonegawa@icluna.kobe-u.ac.jp} and Makoto K{\sevenrm ABURAGI}$^{1,}$\footnote{$^{\ddagger\ddagger}$}{kabu@icluna.kobe-u.ac.jp}}

\vskip 12pt

\centerline{\it Department of Physics, Faculty of Science,
Kobe University,}

\centerline{\it Rokkodai, Kobe 657}

\centerline{$^{1}$\it Department of Informatics, Faculty of
Cross-Cultural Studies,}

\centerline{\it Kobe University, Tsurukabuto, Nada, Kobe 657}

\vskip 12pt

\centerline{(Received May 11, 1995)}

\vskip 12pt

\parindent=1.5pc
Low-lying excited states as well as the ground state of the spin-1
antiferromagnetic Heisenberg chain with a spin-1/2 impurity are studied
by means of a method of numerical diagonalization.  The isotropic
nearest-neighbor exchange coupling plus the uniaxial single-ion-type
anisotropy energy is assumed for the host-system Hamiltonian.   The energy
differences between the ground state and the low-lying excited states, which
appear in the Haldane gap, are estimated in the thermodynamic limit.  The
results are used to analyze the electron-spin-resonance experimental data on
Ni(C$_2$H$_8$N$_2$)$_2$NO$_2$(ClO$_4$), abbreviated NENP, containing a small
amount of spin-1/2 Cu$^{2+}$ impurities.  It is found that in this system, the
impurity-host $\bigl($Cu$^{2+}$-Ni$^{2+}$$\bigr)$ coupling is ferromagnetic
and its magnitude is about 5$\,$\% of the magnitude of the host-host
$\bigl($Ni$^{2+}$-Ni$^{2+}$$\bigr)$ coupling.

\vskip 12pt

\noindent
KEYWORDS:

\noindent
spin-1 antiferromagnetic Heisenberg chain, spin-1/2 impurity, method of
numerical diagonalization, spin-1/2 degrees of freedom, NENP:Cu$^{2+}$ system

\vfill\eject

\hsize=6.46truein
\vsize=8.89truein
\hoffset=-0.43truein
\voffset=0.01truein

\baselineskip=16.7pt

\noindent
{\bf\S1.~Introduction}

\parindent=1.5pc
In the past years a good deal of work has been devoted to the study of the
ground and low-lying excited states of the spin-1 antiferromagnetic
Heisenberg chain.  This is mainly motivated by Haldane's prediction$^{1)}$
that the ground state of the integer-spin case, in contrast to that of the
half-integer-spin case, is a massive state characterized by a finite energy
gap in the excitation spectrum and by an exponential decay of the two-spin
correlation functions.$^{2)}$  In particular, there has recently been
considerable interest in the spin-1/2 degrees of freedom at the edges of the
chain with open boundary conditions.  The spin-1/2 degrees of freedom are well
described by the valence-bond-solid (VBS) picture,$^{3)}$ and lead to the
fourfold degeneracy of the ground state in the limit of infinite lattice since
the effective coupling between them vanishes in this limit.  This degeneracy
has been originally found in the so-called AKLT model$^{3)}$ with open
boundary conditions, which is an exactly solvable model including biquadratic
as well as bilinear exchange interactions.  Later, Kennedy$^{4)}$ has shown
that the ground state of the usual open Heisenberg chain with only bilinear
interactions has the same property.  The present authors and Harada$^{5)}$ have
investigated the low-lying states of the Heisenberg chain with an impurity
bond, which plays a mediating role between the open and periodic chains, in
terms of domain-wall excitations.  They have shown that a coupling between the
spin-1/2 degrees of freedom at the edges via the antiferromagnetic impurity
bond brings about the massive triplet mode in the Haldane gap and that this
massive mode merges the bottom of the energy continuum in the limit of the
periodic chain.  Recently, an appearance of the massive mode in the Haldane gap
has also been discussed by S$\phi$rensen and Affleck.$^{6)}$

\parindent=1.5pc
Hagiwara, Katsumata, Affleck, Halperin, and Renard$^{7\hbox{-}9)}$ have
performed the electron-spin-resonance (ESR) experiment on
Ni(C$_2$H$_8$N$_2$)$_2$- NO$_2$(ClO$_4$), abbreviated NENP, containing a small
amount of spin-1/2 Cu$^{2+}$ impurities, and have given the first experimental
evidence for the existence of the spin-1/2 degrees of freedom at the sites
neighboring the impurity spin.  They have successfully analyzed their
experimental results by using the phenomenological Hamiltonian based on the
VBS picture,$^{3)}$ being composed of three spin-1/2 spins, one of which is 
the Cu$^{2+}$ spin and the other two are the spins corresponding to the
spin-1/2 degrees of freedom.  However, they have not discussed the origin of
the anisotropy of effective exchange interactions between the former and
latter spins, which they have assumed in their analysis.  The spin-1/2 degrees
of freedom at the edges of the open spin-1 antiferromagnetic chain have also
been observed in NENP doped with nonmagnetic ions such as Zn$^{2+}$, Cd$^{2+}$,
and Hg$^{2+}$,$^{10,11)}$ and also even in ^^ pure' NENP in which one may find
finite magnetic chains broken from infinite chains by, for example, crystal
defects.$^{12)}$  It has recently been shown$^{13)}$ that the low-temperature
behavior of the heat capacities and susceptibilities in the NENP:Cu$^{2+}$
system is well described by Hagiwara {\it et al.}'s phenomenological
Hamiltonian.

\parindent=1.5pc
On the other hand, using a quantum Monte Carlo method, Miyashita and
Yamamoto$^{14)}$ have studied the open spin-1 antiferromagnetic chain.  They
have calculated the magnetic moment at each site for the lowest energy state
within the $M_{\rm oc}\!=\!1$ subspace, $M_{\rm oc}$ being the $z$-component
of the total spin for the open chain.  According to their results, the one
half of the total magnetization $M_{\rm oc}\!=\!1$ is localized in each edge
of the chain, and the magnitude of the magnetic moment, which takes a maximum
value at each edge, decays exponentially towards the center of the chain with
the decay constant of about 6 lattice spacings.  Similar results have also been
obtained by White$^{15)}$ by the density-matrix renormalization-group method.
Very recently, Yamamoto and Miyashita$^{16)}$ have extended their
calculation$^{14)}$ to the case of the presence of the uniaxial single-ion-type
anisotropy to show that the spin-1/2 degrees of freedom appear also in this
case, as far as the magnitude of the anisotropy is small enough for the ground
state to be the Haldane state.$^{1)}$

\parindent=1.5pc
In recent papers$^{17,18)}$ the present authors have analytically investigated
the equivalence between the above-mentioned Hagiwara {\it et al.}'s
phenomenological Hamiltonian and more realistic Hamiltonians for the
NENP:Cu$^{2+}$ system.  Then, they have clearly shown that the single-ion-type
anisotropy terms in the latter Hamiltonians play an essential role in
explaining the anisotropy of the effective exchange interactions in the former
Hamiltonian.  This result also confirms the concept of the spin-1/2
degrees of freedom at the edges of the open spin-1 chain.$^{19)}$

\parindent=1.5pc
In the present paper, we aim at discussing quantitatively the properties
of the low-lying excited states as well as the ground state of the spin-1
antiferromagnetic Heisenberg chain with a spin-1/2 impurity.  We express
the Hamiltonian describing this system as
$$ \eqalignno{
   {\cal H} = {\cal H}_{\rm oc}& + {\cal H}'\,,           & (1{\rm a})     \cr
   {\cal H}_{\rm oc}& = J \Bigg\{\sum_{\ell=1}^{N-1}
                            \vecS_\ell \cdot \vecS_{\ell+1}
                      + d \sum_{\ell=1}^{N}\bigl(S_\ell^z\bigr)^2 \Bigg\}
                                 \qquad    (J>0)\,,       & (1{\rm b})     \cr
   {\cal H}'& =
         J'\Bigl(\vecs_0 \cdot \vecS_1 + \vecs_0 \cdot \vecS_N\Bigr)\,,
                                                          & (1{\rm c})     \cr}
$$
where $\vecs_{\rm 0}$ is the spin-1/2 operator of the impurity (Cu$^{2+}$)
spin; $\vecS_\ell$ is the\break
\noindent
spin-1 operator of the host (Ni$^{2+}$) spin; $N$ is the number of host
spins; $J$ and $J'$ are, respectively, the host-host and impurity-host
exchange constants; $d$ is the uniaxial single-ion-type anisotropy constant in
the host system.  Thus, ${\cal H}_{\rm oc}$ and ${\cal H}'$ represent,
respectively, the Hamiltonian for the host system with open boundary conditions
and that for the impurity-host exchange coupling.  It is noted that the
Hamiltonian ${\cal H}$ represents a special (isotropic exchange coupling)
case of the Hamiltonian which we have discussed in ref.$\,$17 as the realistic
one for the NENP:Cu$^{2+}$ system.

\parindent=1.5pc
For the above purpose we employ a method of numerical diagonalization by the
Lancz\"os technique.  We calculate, for various values of $d$ and $J'/J$, the
energies of the ground and low-lying excited states for finite-$N(=\!5$, $7$,
$\cdots$, $17)$ systems.  Then, we extrapolate these finite-$N$ results to the
limit of $N\!\to\!\infty$.  Using the results of the extrapolation, we analyze
the ESR experimental data on the NENP:Cu$^{2+}$ system$^{8,9)}$ to
determine the sign and magnitude of the impurity-host exchange constant $J'$
in this system.

\parindent=1.5pc
In the next section (\S2) the results of the numerical calculation are
presented and discussed.  The analysis of the experimental data is made in
\S3.  Section 4 is devoted to a summary.  In the Appendix, the
expressions for the energies of the low-lying states of the Hamiltonian
${\cal H}$ which we have obtained in ref.$\,$17 are summarized.  We use these
in the discussions in \S2.

\vfill\eject

\hsize=6.56truein
\vsize=8.93truein
\hoffset=-0.43truein
\voffset=0.01truein

\noindent
{\bf\S2.~Results of the Numerical Calculation}

\parindent=1.5pc
In order to calculate the energies of the ground and low-lying excited states
of the system, we first diagonalize numerically, by means of the Lancz\"os
technique, the Hamiltonian ${\cal H}$ for finite $N(=\!5$, $7$, $\cdots$,
$17)$ systems within the subspace determined by the value
$M\!=\!s_0^z\!+\!\sum_{\ell=1}^N\!S_\ell^z$.  In the diagonalization we employ
the computer program package KOBEPACK/I Version 1.0 coded by one of the
present authors (M.~K.).  (The algorithm used in this program package is
discussed in ref.$\,$20.)  Then, we extrapolate the finite-$N$ results
to the limit of $N\!\to\!\infty$.  As shown later, the finite-$N$ results up
to $N\!=\!17$ are sufficient to estimate satisfactorily well the limiting\break
\noindent
($N\!\to\!\infty$) values of the energies at least for the parameter region
which we consider.

\parindent=1.5pc
In the present numerical calculation, we choose the following five values for
$d$:~$d\!=\!0.0$, $0.1$, $0.2$, $0.3$, and $0.4$.  For these $d$'s, the
ground state of the Hamiltonian ${\cal H}_{\rm oc}$ is the Haldane
state.$^{21)}$  We consider only the ground state and the low-lying excited
states which appear in the Haldane gap.  As can be seen from the analytical
result obtained in ref.$\,$17, these states are those with the lowest three
energy eigenvalues within the $M\!=\!1/2$ subspace and that with the lowest
energy eigenvalue within the $M\!=\!3/2$ subspace.  In the following the four
eigenvalues for a given value of $N$ are represented by $\epsilon_{(M,P)}(N)$,
where the different states are denoted by $(M,\,P)$ with\break
\noindent
$P(=\!+$, $-)$ which
stands for the parity with respect to reflection around the impurity spin; we
have two different $(1/2,$ $+)$ states and the prime is added to the state
with the larger eigenvalue to distinguish between these two states.

\parindent=1.5pc
According to our calculation for $N\!=\!5$, $7$, $\cdots$, $17$, the $(1/2,+)$
state is always the ground state of the finite chain.  We define the energy
difference between this state and the other three states as
$$
    \Delta_{(M,P)}(N)
           = \epsilon_{(M,P)}(N) - \epsilon_{(1/2,+)}(N)\,,      \eqno (2)
$$
where $(M,\,P)\!=\!{(1/2,-)}$, $(1/2,+)'$, and $(3/2,+)$.  Estimating
the limiting value $\Delta_{(M,P)}(\infty)$ of $\Delta_{(M,P)}(N)$, we make a
least-squares fit of $\Delta_{(M,P)}(5)$, $\Delta_{(M,P)}(7)$, $\cdots$,
$\Delta_{(M,P)}(17)$ to the following formula:
$$
  \Delta_{(M,P)}(N) = \Delta_{(M,P)}(\infty) + A \exp(-N/\xi)\, N^{-p}\,,
                                                                 \eqno (3)
$$
where $A$, $\xi$, and $p$ are constants which take different values depending
on the three $(M,\,P)$ states.  This procedure leads to the fruitful limiting
results when $0.02\lsim|J'/J|$ $\lsim0.20$ for $d\!=\!0.0$ and when
$0.04\lsim|J'/J|$ $\lsim0.20$ for $d\!>\!0.0$.

\parindent=1.5pc
Using $\Delta_{(M,P)}(\infty)$, we also define, for convenience, the following
quantities:
$$ \!\!
    {\tilde\Delta}_{(1/2,-)}
        = \cases{\Delta_{(1/2,-)}(\infty) & \quad for $J'>0\,$,       \cr
                 \Delta_{(1/2,-)}(\infty) - \Delta_{(1/2,+)'}(\infty)
                                          & \quad for $J'<0\,$,       \cr}
                                                                      \eqno(4)
$$
$$
    {\tilde\Delta}_{(1/2,+)'}
        = \cases{\Delta_{(1/2,+)'}(\infty)  
                   & \qquad\qquad\qquad$\,$ for $J'>0\,$,          \cr
                 -\Delta_{(1/2,+)'}(\infty) 
                   & \qquad\qquad\qquad$\,$ for $J'<0\,$.          \cr}
                                                                     \eqno (5)
$$
$$ \!\!
    {\tilde\Delta}_{(3/2,+)}
        = \cases{\Delta_{(3/2,+)}(\infty) & \quad for $J'>0\,$,    \cr
                 \Delta_{(3/2,+)}(\infty) - \Delta_{(1/2,+)'}(\infty)
                                          & \quad for $J'<0\,$,    \cr}
                                                                     \eqno (6)
$$
It should be noted that
${\tilde\Delta}_{(1/2,-)}$, ${\tilde\Delta}_{(1/2,+)'}$ and
${\tilde\Delta}_{(3/2,+)}$ are the quantities which correspond, respectively,
to $\Delta_{\rm s}(\pm{1/2},\,{1/2})$, $\Delta_{\rm t}(\pm{1/2},\,{3/2})$, and
$\Delta_{\rm t}(\pm{3/2},\,{3/2})$ discussed in the Appendix.

\vskip 8.35pt

\noindent
{\it 2.1~~The case of \hbox{$d\!=\!0.0$}}

\parindent=1.5pc
Let us first discuss the case of $d\!=\!0.0$.  Our calculation shows that
in this case, $\epsilon_{(1/2,+)'}(N)\!=\!\epsilon_{(3/2,+)}(N)$
$\bigl($i.e., $\Delta_{(1/2,+)'}(N)\!=\!\Delta_{(3/2,+)}(N)\bigr)$ holds when
$J'\!>\!0$ and $\epsilon_{(3/2,+)}(N)\!=\!\epsilon_{(1/2,+)}(N)$
$\bigl($i.e., $\Delta_{(3/2,+)}(N)\!=\!0\bigr)$  holds when $J'\!<\!0$.  Thus,
the ground state in the $J'\!<\!0$ case is a quartet, in contrast to the fact
that the ground state in the $J'\!>\!0$ case is a doublet.

\parindent=1.5pc
Tables~\uppercase\expandafter{\romannumeral1}(a),
\uppercase\expandafter{\romannumeral1}(b),
\uppercase\expandafter{\romannumeral1}(c) and
\uppercase\expandafter{\romannumeral1}(d) tabulate the
finite-$N$ values of $\epsilon_{(1/2,+)}(N)$, $\epsilon_{(1/2,-)}(N)$,
$\epsilon_{(1/2,+)'}(N)$, and $\epsilon_{(3/2,+)}(N)$, calculated for various
values of $J'/J$.  The values of these energies for $N\!=\!17$ are plotted as
functions of $J'/J$ in Fig.$\,$1.  We plot the finite-$N$ values of
$\Delta_{(1/2,-)}(N)$ and $\Delta_{(1/2,+)'}(N)$ in Figs.$\,$2(a) and 2(b),
where the limiting values of these quantities estimated
by the procedure discussed above are also shown.

\parindent=1.5pc
The estimated values of ${\tilde\Delta}_{(1/2,-)}$ and
${\tilde\Delta}_{(1/2,+)'}\bigl(=\!{\tilde\Delta}_{(3/2,+)}\bigr)$ are listed
in Table~\uppercase\expandafter{\romannumeral2}, and are plotted versus $J'/J$
in Fig.$\,$3.  In order to obtain an analytical expression for
${\tilde\Delta}_{(M,P)}$ in terms of $J'/J$, we have made a least-squares fit
of these estimated values to a cubic function
$$
 {\tilde\Delta}_{(M,P)}/J 
     = a_{(M,P)}(J'/J) + b_{(M,P)}(J'/J)^2 + c_{(M,P)}(J'/J)^3   \eqno (7)
$$
with the numerical coefficients $a_{(M,P)}$, $b_{(M,P)}$, and $c_{(M,P)}$.  The
results, together with the value $\Delta_{\rm HG}\!=\!0.4105J$$\,^{22,23)}$ of
the Haldane gap, which is defined as the energy difference between the
bottom of the energy continuum and the ground state and is not affected by the
presence of an impurity spin,$^{5)}$ are also shown in Fig.$\,$3.  This figure
should be compared with Fig.$\,$2(b) in ref.$\,$17.  The coefficients
$a_{(M,P)}$ and $b_{(M,P)}$ thus determined numerically are presented in
Table~\uppercase\expandafter{\romannumeral3}.  As is shown in the Appendix,
our analytical study$^{17)}$ has led to the results, $a_{(1/2,-)}\!=\!4/3$
and $a_{(1/2,+)'}\!=\!2$, which are about $5/4$ times as large as
the numerical values.  As far as the ratio $a_{(1/2,-)}/a_{(1/2,+)'}$ is
concerned, however, we have a good agreement between the analytical and
numerical results.  Very recently, S$\phi$rensen and Affleck$^{6)}$ have
obtained the results that $a_{(1/2,-)}\!=\!\alpha$ and
$a_{(1/2,+)'}\!=\!3\alpha/2$ with $\alpha\!\simeq\!1.0640$.  These are in
excellent agreement with our numerical results.

\vskip 8.35pt

\noindent
{\it 2.2~~The case of \hbox{$d\!>\!0.0$}}

\parindent=1.5pc
As mentioned above, we have carried out in this case the numerical calculation
for $d\!=\!0.1$, $0.2$, $0.3$, and $0.4$.  In
Tables~\uppercase\expandafter{\romannumeral4}(a),
\uppercase\expandafter{\romannumeral4}(b),
\uppercase\expandafter{\romannumeral4}(c) and
\uppercase\expandafter{\romannumeral4}(a) we list the
finite-$N$ values of $\epsilon_{(1/2,+)}(N)$, $\epsilon_{(1/2,-)}(N)$,
$\epsilon_{(1/2,+)'}(N)$, and $\epsilon_{(3/2,+)}(N)$ for $d\!=\!0.2$,
calculated for representative values of $J'/J$.  The values of these energies
for $N\!=\!17$ are plotted as functions of $J'/J$ in Fig.$\,$4.  The
finite-$N$ values of $\Delta_{(1/2,-)}(N)$, $\Delta_{(1/2,+)'}(N)$, and
$\Delta_{(3/2,+)}(N)$ for $d\!=\!0.2$ as well as the limiting values of these
quantities estimated by the above procedure are plotted in Figs.$\,$5(a), 5(b)
and 5(c).

\parindent=1.5pc
The estimated values of ${\tilde\Delta}_{(1/2,-)}$,
${\tilde\Delta}_{(1/2,+)'}$, and ${\tilde\Delta}_{(3/2,+)}$ for the
above-mentioned values of $d$ are tabulated in
Table~\uppercase\expandafter{\romannumeral2}, and those for $d\!=\!0.2$ are
plotted versus $J'/J$ in Fig.$\,$6.  In the present case, we have also
obtained the analytical expression for ${\tilde\Delta}_{(M,P)}$ by the same
method as that used in the case of $d\!=\!0.0$, and the results for the
coefficients $a_{(M,P)}$ and $b_{(M,P)}$ are listed in
Table~\uppercase\expandafter{\romannumeral3}.  As we will discuss in \S3, the
Haldane gap $\Delta_{\rm HG}$ for $d\!=\!0.2$ is given by
$\Delta_{\rm HG}\!=\!0.30J$.  This value of $\Delta_{\rm HG}$ as well as the
analytical expression for ${\tilde\Delta}_{(M,P)}$ obtained for $d\!=\!0.2$ is
shown in Fig.$\,$6. This figure should be compared with Fig.$\,$2(a) in
ref.$\,$17.

\vfill\eject

\hsize=6.50truein
\vsize=8.93truein
\hoffset=-0.43truein
\voffset=0.01truein

\noindent
{\bf\S3.~Analysis of the Experimental Data}

\parindent=1.5pc
We now turn to the analysis of the ESR experimental data on the NENP:Cu$^{2+}$
system.$^{8,9)}$  First we discuss the value of $d$ for NENP.
Golinelli, Jolic{\oe}ur, and Lacaze$^{24)}$ have numerically analyzed the
splitting of the triplet state, which lies at the bottom of the energy
continuum in the case of the periodic chain, into singlet and doublet states
due to the uniaxial single-ion-type anisotropy term
$d\sum_{\ell=1}^{N}(S_\ell^z)^2$.  According to their result, the excitation
energy $\Delta_{\rm s}$ of the singlet state measured from the singlet
ground-state energy and the corresponding excitation energy $\Delta_{\rm d}$
of the doublet state are given by
$$ \eqalignno{
     \Delta_{\rm s}& = \bigl(0.41 + 1.41 d\bigr) J\,,   & (8)           \cr
     \Delta_{\rm d}& = \bigl(0.41 - 0.57 d\bigr) J\,,   & (9)           \cr}
$$
for $0.0\!\leq\!d\lsim0.3$.  Golinelli {\it et al.}$^{24)}$ have compared
the inelastic-neutron-scattering (INS) experimental results$^{25)}$
$\Delta_{\rm s}\!=\!2.5\,{\rm meV}$ and
$\Delta_{\rm d}\!=\!1.15\,{\rm meV}$$^{\,26)}$
on NENP with eqs.$\,$(8) and (9) to obtain
$$
   J = 3.75\,{\rm meV} = 43.5\,{\rm K} \qquad\quad {\rm and} 
                                       \qquad\quad d = 0.18\,.      \eqno (10)
$$
On the other hand, if we use the results of magnetization measurements,$^{27)}$
$\Delta_{\rm s}\!=\!27.5\,{\rm K}$ and
$\Delta_{\rm d}\!=\!11.4\,{\rm K}$,$^{\,28)}$ we obtain
$$
   J = 39.1\,{\rm K} \qquad\quad {\rm and} \qquad\quad d = 0.21\,.  \eqno (11)
$$
Thus, we adopt here $d\!=\!0.2$ as the value of $d$ for NENP.

\parindent=1.5pc
Hagiwara {\it et al.}'s ESR experimental results$^{8,9)}$ show that
the ground state and the first, second, and third excited states in the
NENP:Cu$^{2+}$ system are the $M\!=\!\pm{1/2}$, $\pm{3/2}$, $\pm{1/2}$, and
$\pm{1/2}$ states, respectively.  From this together with our analytical
results$^{17)}$ $\big($see Fig.$\,$2 in ref.$\,$17 and eqs.$\,$(A$\cdot$1),
(A$\cdot$2), (A$\cdot$3) and (A$\cdot$4) in the Appendix$\big)$, we can
conclude that $J'$ is ferromagnetic ($J'\!<\!0$).  The energy separations
between the three excited states and the ground state are
$\Delta_1^{\rm exp}\!=\!5.7$GHz, $\Delta_2^{\rm exp}\!=\!25.8$GHz, and
$\Delta_3^{\rm exp}\!=71.7$GHz.  These correspond, respectively, to
$\Delta_{(3/2,+)}(\infty)$,
$\Delta_{(1/2,-)}(\infty)$, and $\Delta_{(1/2,+)'}(\infty)$.  From
Table~\uppercase\expandafter{\romannumeral3} and eqs.$\,$(4-6) we have, for
$-1\!\ll\!J'/J\!<\!0$,
$$ \eqalignno{
   \Delta_{(3/2,+)}(\infty)/J
                    & = - 0.192\,(J'/J) - 0.503\,(J'/J)^2\,,    & (12)   \cr
   \Delta_{(1/2,-)}(\infty)/J
                    & = - 0.592\,(J'/J) + 0.125\,(J'/J)^2\,,    & (13)   \cr
   \Delta_{(1/2,+)'}(\infty)/J
                    & = - 1.618\,(J'/J) - 0.293\,(J'/J)^2\,.    & (14)   \cr}
$$
On the other hand, the Haldane gap $\Delta_{\rm HG}$ for finite $d$ is given
by $\Delta_{\rm d}$, since it should be defined as the energy difference
between the bottom of the energy continuum and the ground state.$^{5)}$  Thus,
$\Delta_{\rm HG}\!=\!0.30J$ for $d\!=\!0.2$, while its experimental value we
use here is
$\Delta_{\rm HG}^{\rm exp}\!=\!1.15\,{\rm meV}\!=\!278\,{\rm GHz}$.$^{25)}$

\parindent=1.5pc
We determine the value of $J'/J$ from
$$
  \Delta_{(1/2,+)'}(\infty) : \Delta_{\rm HG}
          = \Delta_3^{\rm exp} : \Delta_{\rm HG}^{\rm exp}\,,    \eqno (15)
$$
which gives
$$
  J'/J = -0.048\,.                                               \eqno (16)
$$
Then, we have
$$ \eqalignno{
 \Delta_{(3/2,+)}(\infty) : \Delta_{(1/2,-)}(\infty)
                          : \Delta_{(1/2,+)'}&(\infty) : \Delta_{\rm HG}   \cr
                         = &\, 0.03 : 0.10 : 0.26 : 1.00\,, &       (17)   \cr}
$$
and this ratio is in very good agreement with
$$
 \Delta_1^{\rm exp} : \Delta_2^{\rm exp}
                    : \Delta_3^{\rm exp} : \Delta_{\rm HG}^{\rm exp}
                                 = 0.02 : 0.09 : 0.26 :1.00\,.   \eqno (18)
$$

\vfill\eject

\hsize=6.50truein
\vsize=8.93truein
\hoffset=-0.43truein
\voffset=0.01truein

\noindent
{\bf\S4.~Summary}

\parindent=1.5pc
We have investigated the ground and low-lying excited states of the spin-1
antiferromagnetic Heisenberg chain with a spin-1/2 impurity, the Hamiltonian
of which is given by eqs.$\,$(1a-c).  Employing the computer program package
KOBEPACK/I Version 1.0,$^{20)}$ we have calculated, for various values of $d$
and $J'/J$, the energies of these states for finite-$N(=\!5$, $7$, $\cdots$,
$17)$ systems.  We have extrapolated to the limit of $N\!\to\!\infty$ the
finite-$N$ results for the energy difference between the ground state and the
low-lying excited states, assuming the formula given by eq.$\,$(3), to estimate
the values in this limit.  Using these limiting results, we have analyzed
Hagiwara {\it et al.}'s ESR experimental data on the
NENP:Cu$^{2+}$ system.$^{8,9)}$  We have found that in this system, the
magnitude of the impurity-host $\bigl($Cu$^{2+}$-Ni$^{2+}$$\bigr)$ exchange
constant is $0.048$ times as large as that of the host-host
$\bigl($Ni$^{2+}$-Ni$^{2+}$$\bigr)$ exchange constant, the former being
ferromagnetic.

\vskip 35pt

\centerline{\bf Acknowledgments}

The authors would like to thank Drs.~K.~Katsumata and M.~Hagiwara for valuable
discussions.  This work has been supported in part by Grant-in-Aid for
Scientific Research on Priority Areas, ^^ ^^ Computational Physics as a New
Frontier in Condensed Matter Research", from the Ministry of Education,
Science and Culture.

\vfill\eject

\hsize=6.50truein
\vsize=8.93truein
\hoffset=-0.43truein
\voffset=0.01truein

\baselineskip=16.3pt

\noindent
\centerline{\bf Appendix}

\parindent=1.5pc
In ref.$\,$17 we have calculated the energies of four low-lying states of
${\cal H}$, confining ourselves to the Haldane region where the ground state
of ${\cal H}_{\rm oc}$ is the Haldane state, and treating ${\cal H}'$ as a
small perturbation $\bigl(|J'|/J\!\ll\!1\bigr)$.$^{29)}$  In the calculation,
we have employed, as the unperturbed wave functions, the fourfold-degenerate
ground state wave functions of ${\cal H}_{\rm oc}$ obtained in our previous
paper,$^{5)}$ which are expressed in the matrix-product form.$^{30,31)}$  The
calculated energies, measured from the ground state energy $E_{\rm oc}^{(0)}$
$\big($see eq.$\,$(7) in ref.$\,$17$\bigr)$ of ${\cal H}_{\rm oc}$, are given
by
$$ \eqalignno{
     \varepsilon_{\rm t}\biggl(\!\pm{3\over 2},\,{3\over 2}\biggr)
         &= J'\,{4-d\over 6}\,,                       &  ({\rm A}\cdot1)  \cr
     \varepsilon_{\rm t}\biggl(\!\pm{1\over 2},\,{3\over 2}\biggr)
         &= J'\,{4-d\over 12}
            \Biggl\{\sqrt{1+16\,{2+d\over 4-d}}-1
            \Biggr\}\,,                               &  ({\rm A}\cdot2)  \cr
     \varepsilon_{\rm t}\biggl(\!\pm{1\over 2},\,{1\over 2}\biggr)
         &=-J'\,{4-d\over 12}
            \Biggl\{\sqrt{1+16\,{2+d\over 4-d}}+1
            \Biggr\}\,,                               &  ({\rm A}\cdot3)  \cr
     \varepsilon_{\rm s}\biggl(\!\pm{1\over 2},\,{1\over 2}\biggr)
         &= 0\,                                       &  ({\rm A}\cdot4)  \cr}
$$
in the Haldane region $4\!>\!d\!>\!-2$.  Here, we have denoted the
energy of the state belonging to the subspace determined by $M$ as
$\varepsilon_{\rm r}(M,\,S)$ (${\rm r}\!=\!{\rm s}$, ${\rm t}$), where $S$
represents the magnitude of the total spin of the corresponding state in the
isotropic ($d\!=\!0$) case, and the subscripts ${\rm s}$ and ${\rm t}$ show
that the energies are associated with the singlet and triplet states for
${\cal H}_{\rm oc}$, respectively.  If we choose the origin of the
energies $E_{\rm oc}^{(0)}\!+\varepsilon_{\rm t}(\pm{1/2},$ ${1/2})$ and
define $\Delta_{\rm r}(M,\, S)$ (${\rm r}\!=\!{\rm s}$, ${\rm t}$) as
$$  \Delta_{\rm r}(M,\, S)
      = \varepsilon_{\rm r}(M,\, S)
      - \varepsilon_{\rm t}\biggl(\pm{1\over2},\,{1\over2}\biggr)\,,
                                                         \eqno ({\rm A}\cdot5)
$$
then ${\tilde\Delta}_{(1/2,-)}$, ${\tilde\Delta}_{(1/2,+)'}$ and
${\tilde\Delta}_{(3/2,+)}$ correspond, respectively,
to $\Delta_{\rm s}(\pm{1/2},$ ${1/2})$, $\Delta_{\rm t}(\pm{1/2},\,{3/2})$, and
$\Delta_{\rm t}(\pm{3/2},\,{3/2})$.

\vfill\eject

\hsize=6.46truein
\vsize=8.93truein
\hoffset=-0.43truein
\voffset=0.01truein

\baselineskip=16.7pt

\centerline{\bf References}

\parindent=1.5pc

\item{1)} F.~D.~M.~Haldane:~Phys.~Lett.~{\bf 93A} (1983) 464;
Phys.~Rev.~Lett.~{\bf 50} (1983) 1153.

\item{2)} For reviews see, I.~Affleck:~J.~Phys.~Condens.~Matter
{\bf 1} (1989) 3047; M.~Takahashi and T.~Sakai:~{\it Computational Physics as
a New Frontier in Condensed Matter Research}, ed.~H. Takayama, M.~Tsukada,
H.~Shiba, F.~Yonezawa, M.~Imada and Y.~Okabe (Physical Society of Japan, Tokyo,
1995), to be published.

\item{3)} I.~Affleck, T.~Kennedy, E.~H.~Lieb and
H.~Tasaki:~Phys.~Rev.~Lett.~{\bf 59} (1987) 799;~Commun.~Math.~Phys.~{\bf 115}
(1988) 477.

\item{4)} T.~Kennedy:~J.~Phys.~Condens.~Matter {\bf 2} (1990) 5737.

\item{5)} M.~Kaburagi, I.~Harada and T.~Tonegawa:~J.~Phys.~Soc.~Jpn.~{\bf 62}
(1993) 1848.

\item{6)} E.~S.~S$\phi$rensen and I.~Affleck:~Phys.~Rev.~B {\bf 51} (1995)
16115.

\item{7)} M.~Hagiwara, K.~Katsumata, I.~Affleck, B.~I.~Halperin and
J.~P.\break
\noindent
Renard:~Phys.~Rev.~Lett.~{\bf 65} (1990) 3181.

\item{8)} M.~Hagiwara:~Dr.~Thesis, Graduate School of Science, Osaka
University, Toyonaka, Osaka, 1992

\item{9)} K.~Katsumata:~{\it Proc.~Int.~Conf.~on Magnetism, Warsaw, 1994,}
J. Magn.~Magn.~Mater.~{\bf 140-144} (1995) 1595.

\item{10)} S.~H.~Glarum, S.~Geschwind, K.~M.~Lee, M.~L.~Kaplan and
J.~Michel: Phys.~Rev.~Lett.~{\bf 67} (1991) 1614.

\item{11)} Y.~Ajiro:~private communication.

\item{12)} M.~Hagiwara and K.~Katsumata:~J.~Phys.~Soc.~Jpn.~{\bf 61} (1992)
1481.

\item{13)} T.~C.~Kobayashi, H.~Honda, A.~Koda and
K.~Amaya:~J.~Phys.~Soc.~Jpn. {\bf 64} (1995) 2609.

\item{14)} S.~Miyashita and S.~Yamamoto:~Phys.~Rev.~B {\bf 48} (1993) 913.

\item{15)} S.~R.~White:~Phys.~Rev.~Lett.~{\bf 69} (1992)
2863; Phys.~Rev.~B {\bf 48} (1993) 10345.

\item{16)} S.~Yamamoto and S.~Miyashita:~Phys.~Rev.~B {\bf 50} (1994) 6277.

\item{17)} M.~Kaburagi and T.~Tonegawa:~J.~Phys.~Soc.~Jpn. {\bf 63} (1994)
420.

\item{18)} M.~Kaburagi and T.~Tonegawa:~{\it Proc.~4th Int.~Symp.~Research in
High Magnetic Fields, Nijmegen, 1994,} Physica B {\bf 211} (1995) 193.

\item{19)} For a review see, T.~Tonegawa and M.~Kaburagi:~{\it Computational
Phys- ics as a New Frontier in Condensed Matter Research}, ed.~H.~Taka- yama,
M.~Tsukada, H.~Shiba, F.~Yonezawa, M.~Imada and Y.~Okabe (Physical Society of
Japan, Tokyo, 1995), to be published.

\item{20)} M.~Kaburagi, T.~Tonegawa and T.~Nishino:~{\it Computational
Approaches in Condensed Matter Physics}, Springer Proc.~Phys.,
ed.~S.~Miyashita, M.~Imada and H.~Takayama (Springer, Berlin, 1992) p.$\,$179.

\item{21)} R.~Botet, R.~Jullien and M.~Kolb:~Phys.~Rev.~B {\bf 28} (1983) 3914.

\item{22)} S.~R.~White and D.~A.~Huse:~Phys.~Rev.~B {\bf 48} (1993) 3844.

\item{23)} O.~Golinelli, Th.~Jolic{\oe}ur and
R.~Lacaze:~Phys.~Rev.~B {\bf 50} (1994) 3037.

\item{24)} O.~Golinelli, Th.~Jolic{\oe}ur and
R.~Lacaze:~Phys.~Rev.~B {\bf 45} (1992) 9798.

\item{25)} J.~P.~Renard, M.~Verdaguer, L.~P.~Regnault, W.~A.~C.~Erkelens,
J. Rossat-Mignod and W.~G.~Stirling:~Europhys.~Lett.~{\bf 3} (1987)
945;~J.~P. Renard,~M.~Verdaguer, L.~P.~Regnault, W.~A.~C.~Erkelens,
J.~Rossat-Mignod, J.~Ribas, W.~G.~Stirling and
C.~Vettier:~J.~Appl.~Phys.~{\bf 63} (1988) 3538.

\item{26)} Accurate INS measurements$^{25)}$ reveal the small splitting of the
doublet state into two states with the excitation energies $1.05\,{\rm meV}$
and $1.25\,{\rm meV}$, which means that there exists in NENP the orthorhombic
single-ion-type anisotropy in addition to the uniaxial single-ion-type
anisotropy.  Golinelli {\it et al.}$^{24)}$ have used as $\Delta_{\rm d}$ the
arithmetic mean of these excitation energies.

\item{27)} T.~Takeuchi, M.~Ono, H.~Hori, T.~Yosida, A.~Yamagishi and
D.~Date: J.~Phys.~Soc.~Jpn. {\bf 61} (1992) 3255.

\item{28)} Magnetization measurements$^{27)}$ also show the small splitting of
the doublet state into two states, the excitation energies of which are
$10.1\,{\rm K}$ and $12.7\,{\rm K}$.  We use as $\Delta_{\rm d}$ the
arithmetic mean of these excitation energies.

\item{29)} In ref.$\,$17, we have discussed a more general case where both
the host-host and the impurity-hose exchange coupling are anisotropic.

\item{30)} A.~Kl\"umper, A.~Schadschneider and J.~Zittartz:~J.~Phys.~A
{\bf 24} (1991) L955;~Z.~Phys.~B~{\bf 87} (1992) 281.

\item{31)} M.~Fannes, B.~Nachtergaele and
R.~F.~Werner:~Europhys.~Lett.~{\bf 10} (1989)
633;~Commun.~Math.~Phys.~{\bf 144} (1992) 443.

\vfill\eject

\hsize=6.47truein
\vsize=8.89truein
\hoffset=-0.43truein
\voffset=0.01truein

\centerline{\bf Figure Captions}

\vskip 9pt

{\leftskip=1.5pc
\parindent=-1.5pc
Fig.$\,$1.\hbox{~~}Plots versus $J'/J$ of $\epsilon_{(M,P)}(17)$ for
$d\!=\!0.0$.  The open circles show the numerical results for
$\epsilon_{(1/2,+)}(N)$, the crosses show those for $\epsilon_{(1/2,-)}(N)$,
the open triangles show those for $\epsilon_{(1/2,+)'}(N)$, and the solid
circles show those for $\epsilon_{(3/2,+)}(N)$.  The solid lines are the guide
to the eye. \par}

\vskip 9pt

{\leftskip=1.5pc
\parindent=-1.5pc
Fig.$\,$2.\hbox{~~}Plots versus $J'/J$ of $\Delta_{(M,P)}(N)$ for
$d\!=\!0.0$:~(a)~$\Delta_{(1/2,-)}(N)$ and (b)~$\Delta_{(1/2,+)'}(N)$.
The open circles show the numerical results for finite-$N$ systems;~$N\!=\!5$,
$7$, $\cdots$, $17$ from the topmost circle to the lowest circle for each
value of $J'/J$.  The solid lines are the guide to the eye.  The solid squares
show the limiting ($N\!\to\!\infty$) results estimated by the procedure
discussed in the test. \par}

\vskip 9pt

{\leftskip=1.5pc
\parindent=-1.5pc
Fig.$\,$3.\hbox{~~}Plots versus $J'/J$ of ${\tilde\Delta}_{(M,P)}(N)$ for
$d\!=\!0.0$.  The crosses show the estimated values of
${\tilde\Delta}_{(1/2,-)}$ and the open triangles show those of
${\tilde\Delta}_{(1/2,+)'}\bigl(=\!{\tilde\Delta}_{(3/2,+)}\bigr)$.  The solid
lines are least-squares fits to eq.$\,$(7).  The dashed line represents the
bottom of the energy continuum. \par}

\vskip 9pt

{\leftskip=1.5pc
\parindent=-1.5pc
Fig.$\,$4.\hbox{~~}Plots versus $J'/J$ of $\epsilon_{(M,P)}(17)$ for
$d\!=\!0.2$.  The open circles show the numerical results for
$\epsilon_{(1/2,+)}(N)$, the crosses show those for $\epsilon_{(1/2,-)}(N)$,
the open triangles show those for $\epsilon_{(1/2,+)'}(N)$, and the solid
circles show those for $\epsilon_{(3/2,+)}(N)$.  The solid lines are the guide
to the eye. \par}

\vskip 9pt

{\leftskip=1.5pc
\parindent=-1.5pc
Fig.$\,$5.\hbox{~~}Plots versus $J'/J$ of $\Delta_{(M,P)}(N)$ for
$d\!=\!0.2$:~(a)~$\Delta_{(1/2,-)}(N)$,\break
\noindent
(b)~$\Delta_{(1/2,+)'}(N)$, and
(c)~$\Delta_{(3/2,+)}(N)$.  The open circles show the numerical results for
finite-$N$ systems;~$N\!=\!5$, $7$, $\cdots$, $17$ from the topmost circle to
the lowest circle for each value of $J'/J$.  The solid lines are the guide to
the eye.  The solid squares show the limiting ($N\!\to\!\infty$) results
estimated by the procedure discussed in the test. \par}

\vskip 9pt

{\leftskip=1.5pc
\parindent=-1.5pc
Fig.$\,$6.\hbox{~~}Plots versus $J'/J$ of ${\tilde\Delta}_{(M,P)}(N)$ for
$d\!=\!0.2$.  The crosses show the estimated values of
${\tilde\Delta}_{(1/2,-)}$, the open triangles show those of
${\tilde\Delta}_{(1/2,+)'}$, and the solid circles show those of
${\tilde\Delta}_{(3/2,+)}$.  The solid lines are least-squares fits to
eq.$\,$(7).  The dashed line represents the bottom of the energy continuum.
\par}

\bye